# Experimental investigation of pulsed entangled photons and photonic quantum channels

Yoshihiro Nambu[a,c], Koji. Usami[b,c], Akihisa Tomita[a,c,d], Satoshi Ishizaka[a,c], Tohya Hiroshima[a,c], Yoshiyuki Tsuda[d], Keiji Matsumoto[d], and Kazuo Nakamura[a,b,c] *

[a] Fundamental Research Laboratories, NEC Corporation, 34, Miyukigaoka, Tsukuba, Ibaraki 305-8501, Japan

[b] Department of Material Science and Engineering, Tokyo Institute of Technology, 4259, Nagatsuta-chou, Midori-ku, Yokohama, Kanagawa, 226-0026, Japan

[c] CREST, Japan Science and Technology Corporation (JST)

[d] ERATO, Japan Science and Technology Corporation (JST), 5-28-3, Hongo, Bunkyo-ku, Tokyo, 113-0033, Japan

## ABSTRACT

The development of key devices and systems in quantum information technology, such as entangled particle sources, quantum gates and quantum cryptographic systems, requires a reliable and well-established method for characterizing how well the devices or systems work. We report our recent work on experimental characterization of pulsed entangled photonic states and photonic quantum channels, using the methods of state and process tomography. By using state tomography, we could reliably evaluate the states generated from a two-photon source under development and develop a highly entangled pulsed photon source. We are also devoted to characterization of single-qubit and two-qubit photonic quantum channels. Characterization of typical single-qubit decoherence channels has been demonstrated using process tomography. Characterization of two-qubit channels, such as classically correlated channels and quantum mechanically correlated channels is under investigation. These characterization techniques for quantum states and quantum processes will be useful for developing photonic quantum devices and for improving their performances.

## 1. INTRODUCTION

In recent years, considerable efforts have been devoted to develop the physical devices realizing various concepts proposed in a quantum information technology, such as quantum communication channels, quantum sources for quantum information carriers, quantum gates that perform quantum operations on a single qubit or two qubits, and quantum measurement devices. In these developments, the evaluation of the device under development is of great importance. For example, suppose that one is developing a quantum source for information carriers that is designed to prepare a particular initial quantum state. We usually evaluate the prepared state by changing certain controllable parameters of the device to find the optimum operating conditions under which the desired initial state can be achieved. For the other example, suppose that one is also attempting to implement a quantum gate, which is described by the joint unitary operation over two qubits. Here, it is important to evaluate experimentally the performance of the gate and use the results to refine it.

An important feature of these quantum devices is that they are open systems; that is, they are usually coupled to an environment with uncontrollable degree of freedom. As a result, the devices inevitably encounter some noise, or in other words, decoherence. Much of the recent experimental effort has been devoted to reducing decoherence by decoupling the relevant degree of the freedom of the qubits from the degrees of freedom of the environment, thereby realizing an ideal quantum device. Thus an experimental technique that can reveal the system performance as well as the detailed information about system-environment interaction would be valuable for the development of the device.

Y.N.: E-mail: y-nambu@ah.jp.nec.com

Over the past few years, a lot of authors have paid great attention to the problem of characterizing quantum states[1-3], physical devices[4-7], and quantum measurement.[8]

In this paper, we present our most recent experimental work on the characterization of the quantum states and quantum channels and their applications. First, we present the characterization of the quantum states of entangled qubit pairs realized using the polarization degrees of freedom of a pair of photons created in a parametric down-conversion experiment. By using quantum state tomography, we have been able to reconstruct the polarization state of the photon pairs generated from the two-photon source under development at our laboratory. This information has given us the physical origin of decoherence in the generated entangled state, and with this knowledge, we have successfully eliminated the decoherence effect and recovered a highly entangled state The paper also presents preliminary experimental results on the characterization of the quantum channels that represent certain quantum operations on a single qubit realized using the polarization degrees of freedom of a single photon. Our main concern is the characterization of typical single-qubit decoherence channels by using process tomography. Characterization of two-qubit channels, such as the classically correlated channel and the quantum correlated channel is currently under investigation.

## 2. CHARACTERIZATION OF QUANTUM STATES

Photon pairs generated in the process of spontaneous parametric down-conversion (SPDC) have been an effective and convenient source of two-particle entangled states and have been used in tests of the foundations of quantum mechanics as well as quantum information technologies such as quantum cryptography and quantum teleportation. In such applications, a pulsed source of entangled photon pairs is particularly useful because the times of emission are known to the users. We have characterized the state created from a pulsed entangled photon source by using quantum state tomography, and have carefully examined the dependence of this state on the various experimental parameters. We have found that a highly entangled photon state can be obtained by choosing the appropriate experimental parameters.[9]

Our experimental setup is schematically shown in Fig. 1. Two adjacent, thin, type-I crystals (BBO) whose optic axes are horizontally (H) and vertically (V) oriented, respectively, are pumped by 45°- polarized femtosecond 266-nm pulses. The approximate average power of the pump pulses was 150 mW, and the repetition rate was 82 MHz. Because of the type-I coupling, $H$-polarized photon pairs of 532 nm are generated by the $V$-polarization component of the pump field in the first crystal, and $V$-polarized photon pairs are generated by the $H$-polarization component of the pump field in the second crystal. These two down-conversion processes are equally likely and are coherent with one another[10, 11]. The SPDC was performed under a degenerate and quasi-collinear condition. The signal and idler photons making an angle of 3° with respect to the pump laser beam and having the same wavelength (around 532 nm) were observed through identical interference filters, centered at 532 nm, placed in front of the detectors. The created photon pairs were observed through polarization analyzers, each consisting of a rotatable half- wave plate (HWP) and a quarter-wave plate (QWP) (for 532 nm) followed by a polarizing beam splitter (PBS). After passing through the irises, the photons were collected with 60-mm-focal-length lenses, and directed onto the detectors. The detectors were photomultipliers (PMT) (HAMAMATSU H7421-40) placed about 1.5 m from the crystal, with efficiencies of about 40% at 532 nm and dark count rates of the order of 80 s? ¹. The outputs of the detectors were recorded using a time interval analyzer (YOKOGAWA TA-520), and pulse pairs received within a time window of 7 ns were counted as coincident.

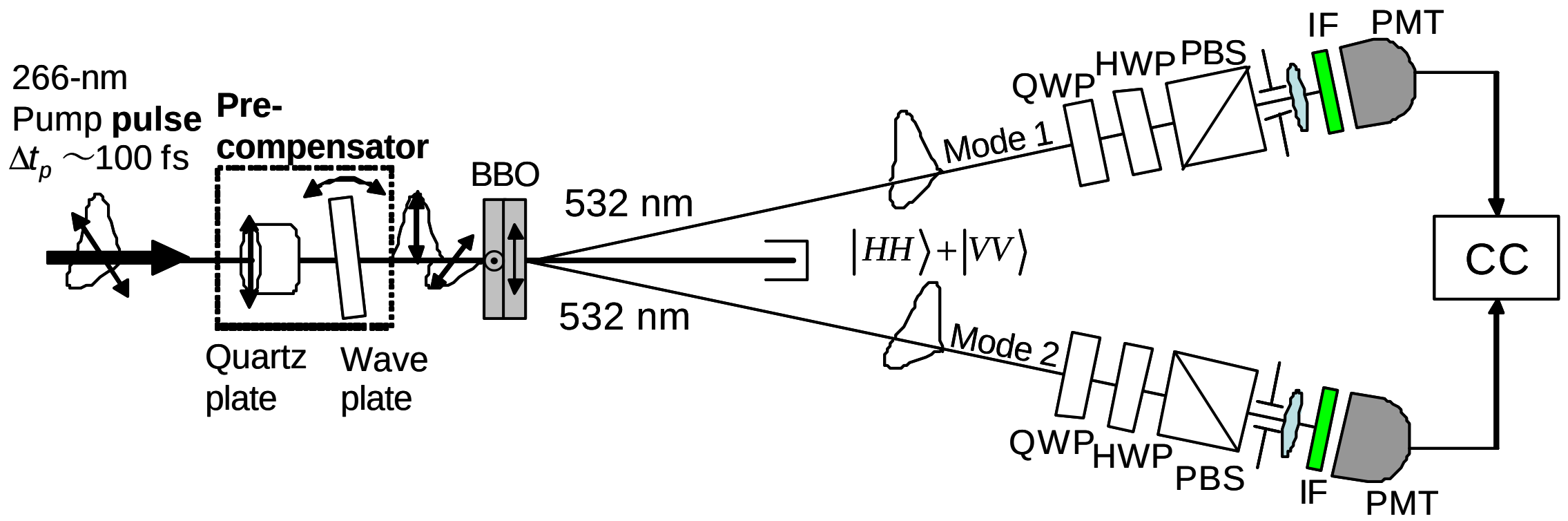

**Figure 1** Experimental setup for generating and characterizing polarization entangled pulsed photons. The pump pulse polarized at 45° irradiates a cascade of two type-I crystals, whose optic axes are orthogonally oriented. A pre-compensator (quartz plate and tiltable wave plate) introduces relative delays between the two-photon amplitudes created in the first and second crystal. CC is the coincidence circuit.

To obtain a truly polarization-entangled state, one must disentangle the polarization degree of freedom from all other degrees of freedom; that is, one must factorize the total state into the product of the polarization-entangled state and those describing the other degrees of freedom. In the present case, effective polarization entanglement requires the suppression of any distinguishing information in the other degrees of freedom that can provide potential information about "which polarization" the emitted photon pair has as well as "which crystal" each pair originated from, since the former and latter are intrinsically correlated. Accordingly, to make the emitted spatial modes for a given pair indistinguishable for the two crystals, we must spatially overlap the down-conversion light cones by using very thin crystals, each having a thickness of about 130 μm. In addition, it is important to eliminate distinguishing space-time information inherent in the two-photon states produced in the ultrashort-pulse-pumped SPDC. Figure 2 schematically illustrates what may happen when a 45°-polarized femtosecond pulse is incident on two cascaded type-I BBO crystals, which has been roughly estimated from the optical characteristics of the BBO[12]. For simplicity, we consider here the case of degenerate collinear SPDC. Because of the combined effects of group velocity dispersion and birefringence in the two crystals, the space-time components of the two-photon state associated with the polarization states $|HH\rangle$ and $|VV\rangle$ are expected to be temporally displaced by approximately 168 fs after the crystals, where $|H\rangle$ ($|V\rangle$) means a single photon linearly polarized along the horizontal (vertical) axis and the first (second) letter corresponds to the signal (idler).[9] Since down-converted photon wavepackets are expected to have widths comparable to that of the pump field (about 150 fs), it is suggested that the space-time components associated with $|HH\rangle$ and $|VV\rangle$ do not overlap in space-time. As a result, "which-polarization" information may be available, in principle, from the arrival time of the photons at the detector.

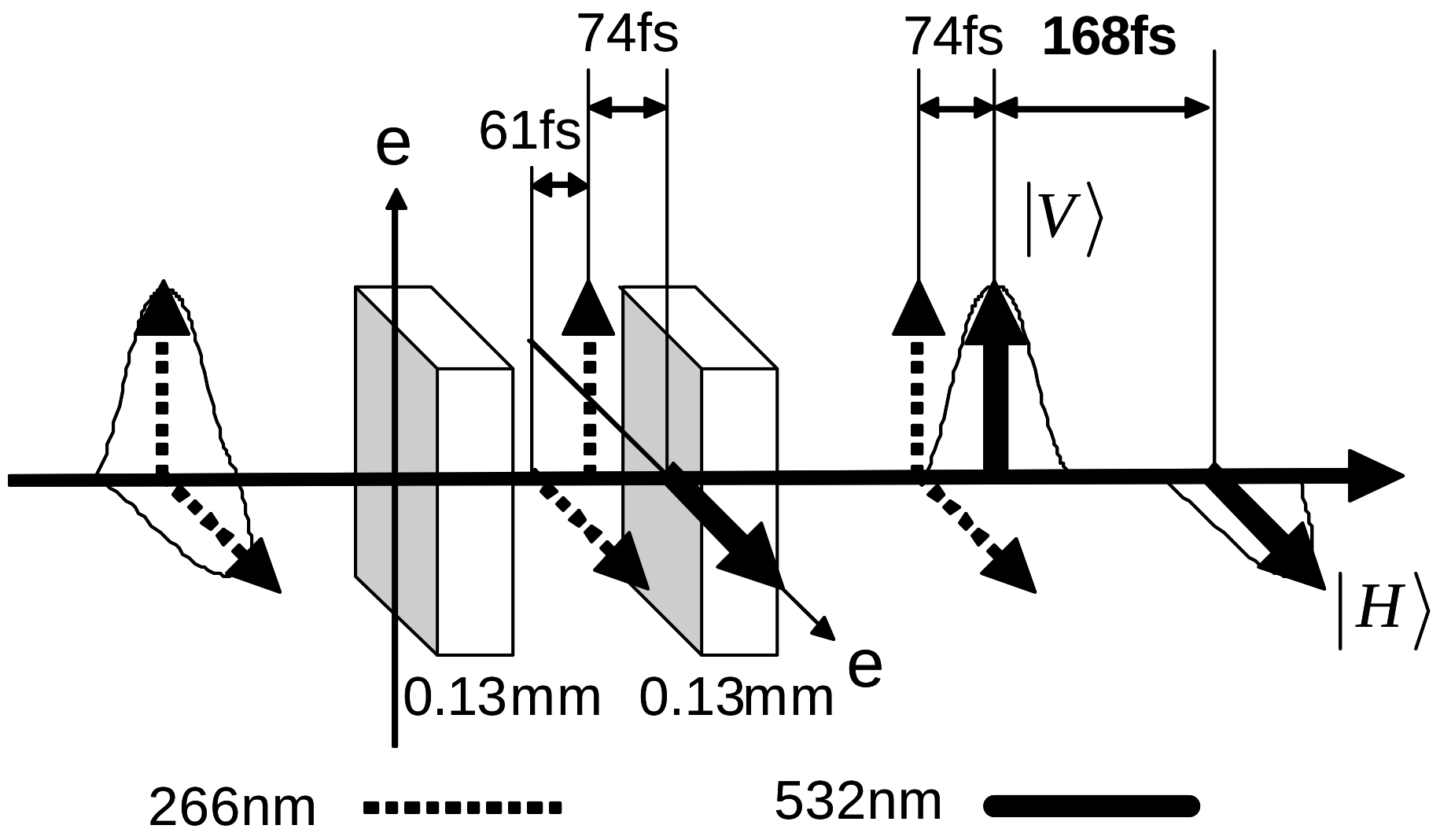

**Figure 2** Schematic diagram of the temporal development of the down-converted photons and the pump .[9]

To eliminate this distinguishing space-time information, a relative delay of $|T|$=350 fs was introduced between the *H*- and *V*-polarization components of the 266-nm pump field by using a polarization-dependent optical delay line inserted before the crystals, which is denoted as the pre-compensator in Fig. 1. It consists of quartz plates, whose optic axes are oriented either vertically or horizontally, and a Bereck-type tiltable polarization compensator. This delay line can compensate for the relative delay between the state associated with $|HH\rangle$ created at the first crystal relative to the state associated with $|VV\rangle$ created at the second crystal, in order to overlap them. The Bereck-type polarization compensator is used to adjust the subwavelength delay. As a result, a two-photon Bell state $|F^{+}\rangle=(|HH\rangle+|VV\rangle)/\sqrt{2}$ can be directly created.

Two sets of experiments were carried out to confirm whether the target state $|F^{+}\rangle$ was successfully prepared. The first experiment was a quantum state tomography[2, 3]. The polarization density matrices were estimated from 16 kinds of joint projection measurements performed on an ensemble of identically prepared photon pairs. These joint measurements consisted of four kinds of projections onto $\{|H\rangle, |V\rangle, |D\rangle, |L\rangle\}$ for each member of a photon pair, where $|D\rangle=(|H\rangle+|V\rangle)/\sqrt{2}$ and $|L\rangle=(|H\rangle+i|V\rangle)/\sqrt{2}$. We used a maximum likelihood calculation[3] to estimate the density matrix. We also calculated the concurrence of the state, which is known to give a good measure of the entanglement of a two-qubit system.[13] The second experiment was a conventional two-photon polarization interference experiment where the projection measurements onto $\{|L\rangle$ or $|R\rangle\}$ ? $|?\rangle$, where $|R\rangle=(|L\rangle)^*$ and $|?\rangle=\cos 2?|H\rangle+i\sin 2?|V\rangle$, were performed on each member of the photon pair. The visibility of the observed interference pattern gives another convenient measure of the degree of entanglement.

Figure 3 summarizes the main results of the experiment. This experiment used interference filters having a full-width half maximum (FWHM) bandwidth of 8 nm, in front of the detectors. The middle traces indicate the estimated polarization density matrices when (a) no relative delay, (b) a relative delay of 135 fs, or (c) a relative delay of 231 fs was introduced between the *H*- and *V*-polarization components of the 266-nm pump field. As shown in the figures, we could successfully prepare a mixture of states approximately described by $?(v)= v|F^{+}\rangle\langle F^{+}| + ((1-v)/2)(|HH\rangle\langle HH|+|VV\rangle\langle VV|)$. These states should exhibit interference in the coincidence rates in proportion to $P(?)=(1/4)(1\pm v\sin 4?)$. The lower traces of Fig. 3 show the coincidence rates as a function of the HWP angle in mode 2 and the evaluated visibilities associated with the states in the middle traces of Fig. 3(a)-(c). Note that no background was subtracted in these results, so that the evaluated visibilities give the lower limits of *v*. The experiments were repeated for various values of delay *T* introduced in the *H*- and *V*-polarization components of the pump field. The evaluated concurrence *C* and visibility *v* of the polarization interference are plotted in the upper trace of Fig. 3 as a function of relative delay *T*. This figure clearly indicates that there is a strong correlation between the concurrence and the visibility. This is quite natural because $C=v$ should hold for the mixed states $?(v)$ given by the above form. The maximum values of the concurrence and visibility were 0.95 and 0.92±0.02, respectively. The broken line is a Gaussian curve fitted to the measured data, and the FWHM width is 237 fs.

To summarize this section, we could develop a quantum source of entangled photon pairs by carefully examining the dependence of the created state on the experimental parameters. The ability to characterize the quantum states will provide us with a powerful tool for developing novel quantum sources for information carriers.

## 3. CHARACTERIZATION OF QUANTUM CHANNELS

The rest of this paper presents recent experimental results on the characterization of the quantum channels. Before presenting the details of our experiment, it will be helpful to introduce the notion of the quantum channel and explain the significance of its characterization when developing physical devices for a quantum information technology. The quantum channel is the most general notion in the field of quantum information technology. Logically, it is a "black box" with inputs and outputs that transforms an arbitrary input state into an output state determined by a transfer function obeying quantum physics. The box may even be coupled to an environment or have other inputs and outputs, which we wish to ignore; i.e., it may be an open quantum system. Any quantum mechanical process can be viewed as a quantum channel. For example, physical objects such as optical fiber, parametric amplifiers, directional couplers, C-Not gates, quantum cloning machines, quantum communication channels, etc., can be viewed as quantum channels.

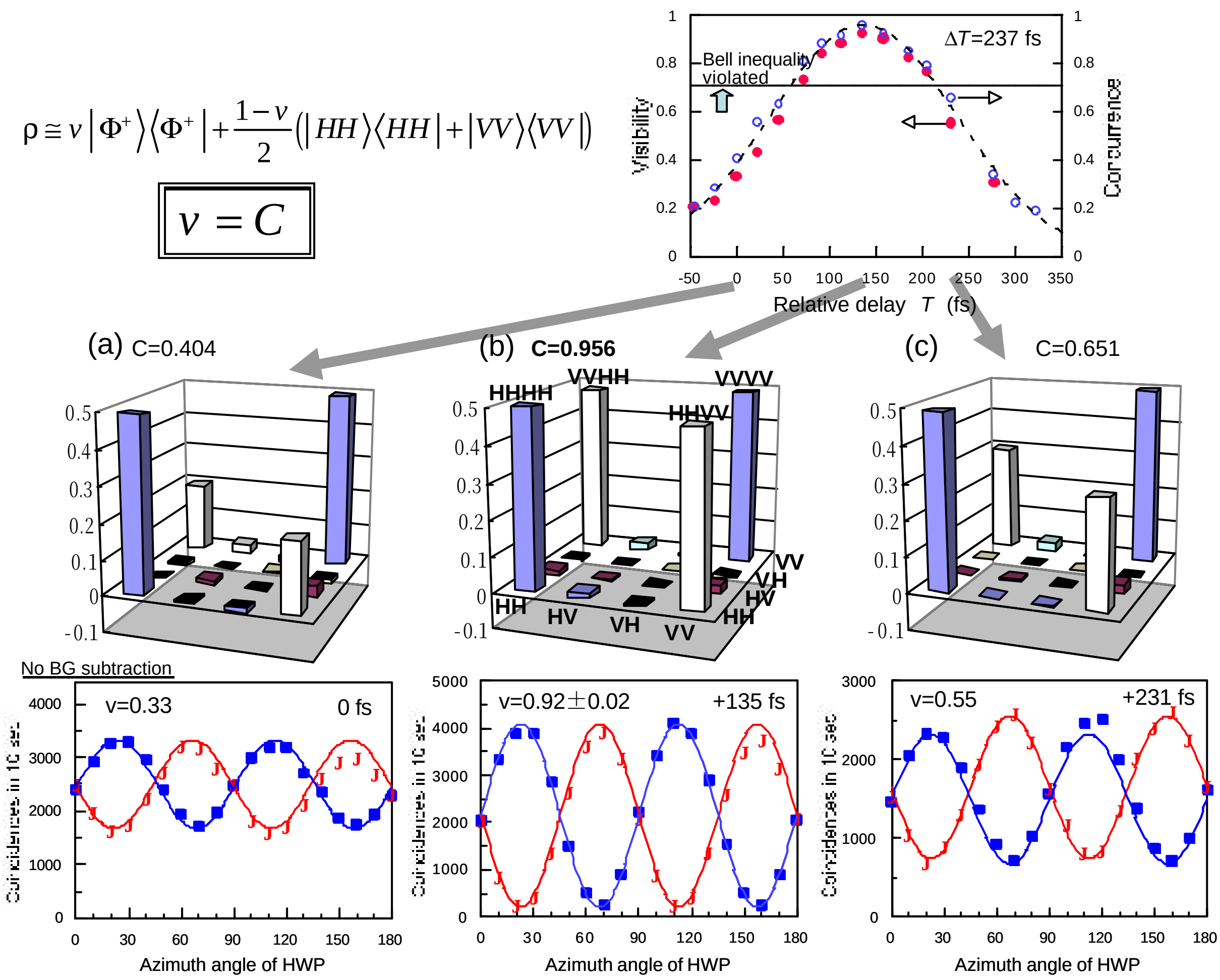

**Figure 3** Summary of the experimental results. The upper trace: Evaluated concurrence $C$ and visibility $v$ are plotted against the relative delay $T$ between two orthogonal components of the pump field. The broken line is the Gaussian fit of the data with 237-fs widths (FWHM). The middle traces: Estimated polarization density matrices for (a) no relative delay, (b) relative delay of 135 fs, and (c) relative delay of 231 fs between the $H$- and $V$-polarization components of the pump field. Only the real parts of the matrices are shown. The contribution of the imaginary parts is negligible. The dark-colored bars show diagonal elements, while the light-colored bars show off-diagonal elements. Estimated values of concurrence are shown as $C$. The lower traces: Polarization correlation experiment. Coincidence counts in 10 s are plotted against the orientation angle of the half-wave plate in mode 2, while the half-wave plate in mode 1 was fixed at ±22.5°, for the states shown in the middle traces. Estimated visibilities are also shown in the figure.

Even quantum information protocols and algorithms can be viewed as quantum channels. For example, a quantum cryptography channel (including an eavesdropper), quantum teleportation[14], or quantum computer executing a certain algorithm may also be viewed as a quantum channel. Therefore, an experimental technique to characterize quantum channel would allow us to investigate how well a physical device or quantum protocol or algorithm implemented in the real world works. It also would allow us to gain important information about the interaction between different systems, typically a system-environment interaction. Such information may be practically useful to improve the performance of a device, protocol or algorithm under development.

From a theoretical point of view, a quantum channel is described in terms of a completely positive linear map $\mathcal{E}$ (CP linear map) that maps an input density matrix $\rho_{in}$ onto an output density matrix $\rho_{out}$: $r_{out}=\mathcal{E}(r_{in})$. Our aim is to characterize the CP map $\mathcal{E}$, given as a "black box", by a sequence of measurements in such a way that it is possible to infer what the output state will be for any given input state. If the trace of the density matrix does not change because of the map $\mathcal{E}$, i.e., tr[$r_{out}$]=tr[$\mathcal{E}(r_{in})$], $\mathcal{E}$ is said to be trace-preserving. In the following, we consider a trace-preserving, CP linear map $\mathcal{E}$. Note that any input density matrix in Hilbert space $\mathcal{H}$ can be written as a linear combination of orthonormal basis $\{L_q\}$ for the space of linear operators on $\mathcal{H}$, namely tr[$L_q^{\dagger} L_{q'}$]=$\delta_{qq'}$,

$$r_{in} = \sum_q l_q L_q ,$$

where $l_q$=tr[$r L_q$] and $\Sigma_q\, l_q=1$. The output state can be then written as a linear combination in the following form:

$$r_{out} = \mathcal{E}(r_{in}) = \sum_q l_q \mathcal{E}(L_q) .$$

This implies that to know the CP map $\mathcal{E}$, it is sufficient to know $\{\mathcal{E}(L_q)\}$. For the input with a state space of $N$ dimensions, we need at least $N^4 - N^2$ kinds of experiments to know $\{\mathcal{E}(L_q)\}$.[4-8] In the following, we will limit our discussion to the case where the input system is a single qubit, realized using the polarization degrees of freedom of a single photon, which is directly related to our experiment presented later.

There is arbitrariness in choosing the orthonormal basis set $\{L_q\}$. We choose the following basis set: $L_0=|H\rangle\langle H|$, $L_1=|H\rangle\langle V|$, $L_2=|V\rangle\langle H|$, $L_3=|V\rangle\langle V|$. As a result, $\mathcal{E}(L_0)$ and $\mathcal{E}(L_3)$ can be obtained simply by measuring the state output from the "black box" associated with the $H$- and $V$-polarized input states, by using state tomography for a qubit. On the other hand, $\mathcal{E}(L_1)$ and $\mathcal{E}(L_2)$ can be obtained by preparing the input states $|H\rangle$, $|V\rangle$, $|D\rangle$, $|L\rangle$ and forming a linear combination of $\mathcal{E}(|H\rangle\langle H|)$, $\mathcal{E}(|V\rangle\langle V|)$, $\mathcal{E}(|D\rangle\langle D|)$, and $\mathcal{E}(|L\rangle\langle L|)$, as follows: [5]

$$\mathcal{E}(\Lambda_1) = \mathcal{E}\left(|H\rangle\langle V|\right) = \mathcal{E}\left(|D\rangle\langle D|\right) + i\mathcal{E}\left(|L\rangle\langle L|\right) - \frac{1+i}{2}\left\{\mathcal{E}\left(|H\rangle\langle H|\right) + \mathcal{E}\left(|V\rangle\langle V|\right)\right\} , \qquad (1)$$

and $\mathcal{E}(L_2)$ is given by taking the complex conjugate of $\mathcal{E}(L_1)$. Thus, it is possible to determine $\{\mathcal{E}(L_q)\}$ from the state tomography of the output state $\mathcal{E}(|H\rangle\langle H|)$, $\mathcal{E}(|V\rangle\langle V|)$, $\mathcal{E}(|D\rangle\langle D|)$, and $\mathcal{E}(|L\rangle\langle L|)$.

Every CP linear map $\mathcal{E}$ has its operator sum representation (or Steinespring decomposition):

$$\mathcal{E}(r) = \sum_i A_i\, r\, A_i^{\dagger} , \qquad (2)$$

where the $A_i$'s are operators on $\mathcal{H}$. The map $\mathcal{E}$ is trace-preserving if and only if $\sum_i A_i^{\dagger} A_i = I$, where $I$ denotes the identity operator on $\mathcal{H}$. However, in some situations, it is advantageous to use another form of the decomposition in order to obtain physical insight from the experimental results:

$$\mathcal{E}(r) = \sum_{i,j=0}^{3} c_{ij}\, s_i\, r\, s_j , \qquad (3)$$

where $\sigma_0 = I$ and $\sigma_i$ ($i$=1,2,3) is the Pauli operator. The matrix $\chi$ is positive Hermitian and completely describes $\mathcal{E}$. There is a simple algebraic relationship between $\{\mathcal{E}(L_q)\}$ and $\chi$-matrix, and one can reconstruct $\chi$-matrix from $\{\mathcal{E}(L_q)\}$ through a simple matrix calculation. Once $\chi$ is obtained, we can calculate $A_i$ by finding the unitary matrix that diagonalizes $\chi$-matrix.[5,8] In the experiment presented below, we determined $\chi$-matrix from a set of experiments that determine $\{\mathcal{E}(L_q)\}$.

Figure 4 shows the experimental setup. It is almost the same as the setup of the previous experiment. Thin type-I crystals (BBO) whose optic axis is $V$-oriented is pumped by $V$-polarized femtosecond 266-nm pulses. Because of the type-I coupling, $H$-polarized photon pairs of 532 nm are generated. The signal and idler photons making an angle of 3° with respect to the pump laser beam and having the same wavelength were observed through the identical 8-nm-width filters, centered at 532 nm, placed in front of the detectors. One of the photon pair was used as the trigger of occurrence of a single photon emission of the other pair, i.e., whether a target qubit is prepared. The created photon was unitary-transformed into the input states $|H\rangle$, $|V\rangle$, $|D\rangle$, and $|L\rangle$ through the combination of rotatable HWP and QWP. The photon then passed through the quantum channel under investigation and was subsequently observed through polarization analyzers (rotatable HWP and QWP (for 532 nm) followed by a PBS). The coincidence counts of the output of the detectors were recorded. Associated output states $\mathcal{E}(|H\rangle\langle H|)$, $\mathcal{E}(|V\rangle\langle V|)$, $\mathcal{E}(|D\rangle\langle D|)$, and $\mathcal{E}(|L\rangle\langle L|)$ were determined by state tomography, and $\chi$-matrix was calculated.

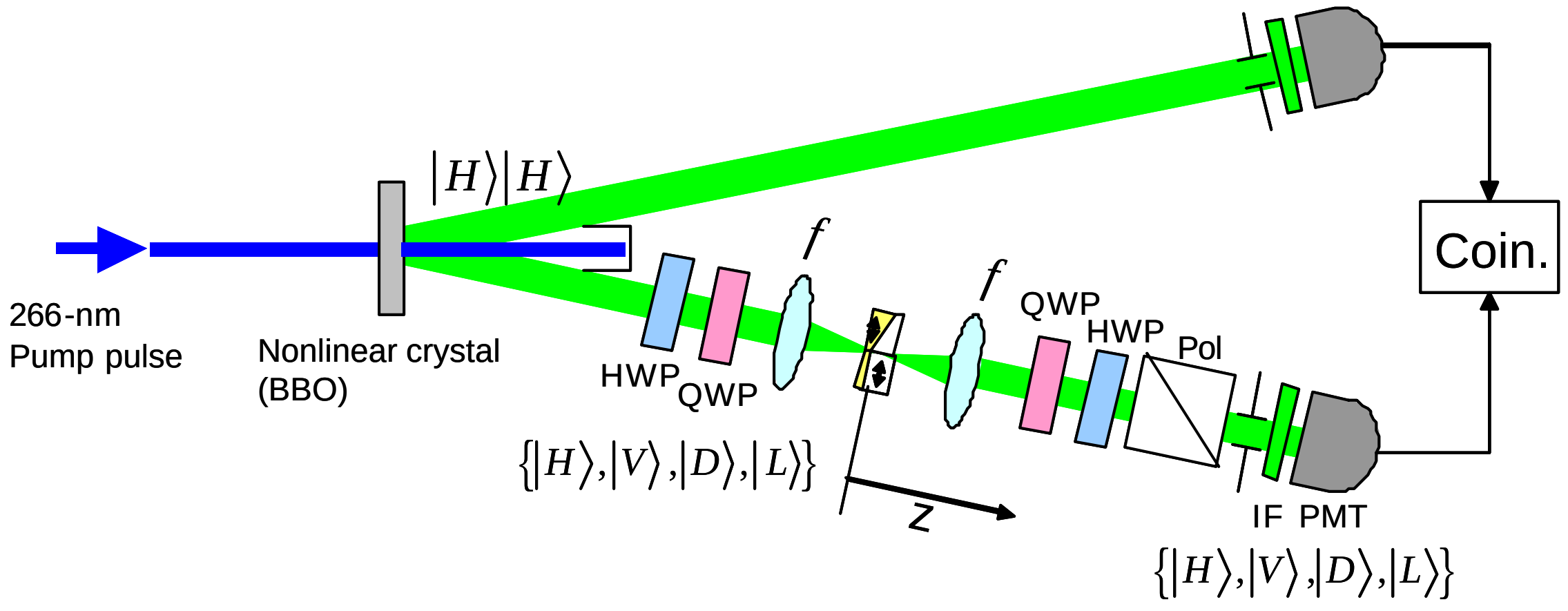

**Figure 4** Experimental setup for characterizing a single-qubit quantum channel, realized using the polarization degrees of freedom of a single photon. A vertically polarized pump pulse produces a horizontally polarized photon pair, one of which is used as a trigger for a single photon event. The other is used to characterize the quantum channel. The quantum channel is a depolarizer (a wedge shaped quartz plate). The photon field is focused on the plane of the depolarizer by using a pair of focusing lenses.

In the following, we show the feasibility of the experiment on the basis of examples of a single qubit channel, that is, a quantum channel that represents a quantum operation on a single qubit. A single qubit channel that is isolated from the environment is a trivial unitary operation. What we are curious about, here, is a single qubit channel that represents an open quantum system. One such channel is a decoherence channel that is unital or bistochastic; i.e., the associated map preserves both the trace and the identity. To realize such a channel, we used a wedge shaped birefringent quartz plate as a depolarization plate. Because the thickness of the birefringent plate depends on the location on the plane of the plate, it acts as a spatially inhomogeneous phase plate for two orthogonal linearly polarized photons. If the depolarizer whose optic axis is parallel to (oriented 45° with respect to) the *H*- or *V*-direction is used as a quantum channel, the photon experiences a random unitary rotation around the z (x) axis in the polarization space. Furthermore, the inhomogeneity the photon field depends on the beam diameter on the plane of the depolarizer plate, so that we can effectively control the randomness by changing the diameter of the photon field. This is achievable by placing identical focusing lenses ($f$=20 mm) before and after the depolarizer and by changing the relative position of two lenses and the depolarizer as shown in Fig. 4. This setup can simulate the channel given by an action of an external random field.[6,17]

At this point, we should consider the underlying assumption of the above discussion. We have assumed that the measurement apparatus is perfect. If the apparatus itself has imperfections, then the characterization of $\mathcal{E}$ would be distorted. To confirm the validity of the perfect apparatus assumption, we characterized a null channel ($\rho_{out}=\rho_{in}$), i.e., a channel with no elements. Figure 5 shows the $\chi$-matrix for the null channel. It indicates that the elements other than $\chi_{00}$ are negligible. This implies that our experimental apparatus is nearly perfect, so the assumption is valid.

We have characterized three typical decoherence channels as an illustration of the process tomography. The first is a quantum channel that preserves quantum states as much as possible, which simulates an imperfect quantum communication channel. Suppose that the photon experiences a random unitary operation only around the x-axis in the polarization space. The operator sum representation of this channel is given by

$$\mathcal{E}(r) = \frac{1+|s|}{2} U_x(0)\, r U_x^\dagger(0) + \frac{1-|s|}{2} U_x(-p)\, r U_x^\dagger(-p)\,. \qquad (4)$$

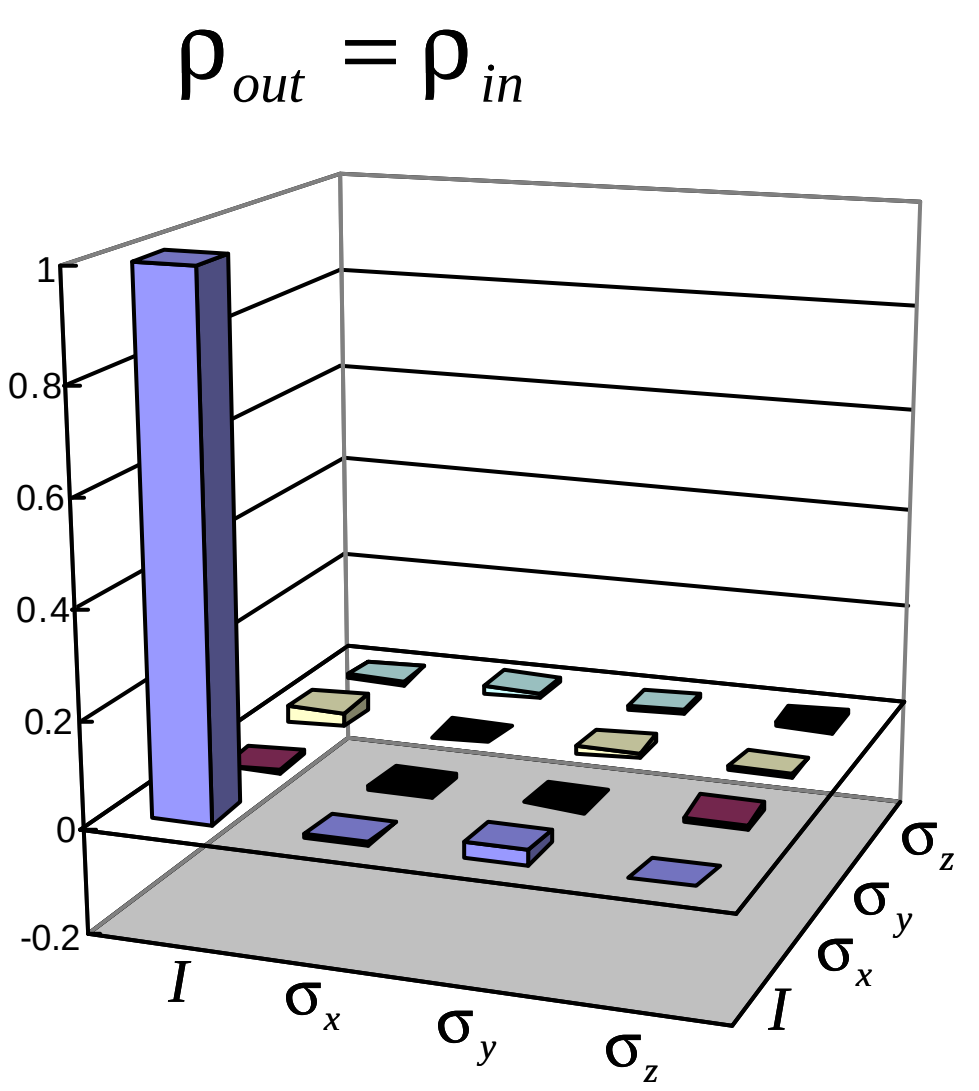

**Figure 5** Evaluated χ-matrix for the null channel ($r_{\text{out}}$=$r_{\text{in}}$). Only the real parts of the matrix elements are shown (the imaginary parts are negligible). The indices of the matrix elements are shown in the figure.

This channel is called a partially bit-flip channel and can be realized by placing a depolarizer whose optic axis is oriented at 45° with respect to the *H*- or *V*-direction for the channel under test. Figure 6 shows the experimental results corresponding to this situation. The upper trace indicates the degree of polarization of the photon at the detector as a function of the relative position *z* of the two lenses and the depolarizer for four input polarization states. The lower traces are χ-matrix for *z*=0, 2, 5, and 8 mm, which are in good agreement with the theoretical χ-matrix:

$$c = \frac{1}{2}\begin{pmatrix} 1+|s| & 0 & \multicolumn{2}{c}{\mathbf{0}} \\ 0 & 1-|s| & & \\ \multicolumn{2}{c}{\mathbf{0}} & \multicolumn{2}{c}{\mathbf{0}} \end{pmatrix}, \qquad (5)$$

associated with various values of *s*. Next, we placed a depolarizer whose optic axis is parallel to the *H*- or *V*-direction after the 45°-oriented depolarizer. If we assume that the randomization process occurs independently in the first and second depolarizer, the associated operator sum representation is given by

$$\mathcal{E}(r) = \left(1-p-q-r\right) r + pU_x(-p)\, rU_x^\dagger(-p) + qU_y(-p)\, rU_y^\dagger(-p) + rU_z(-p)\, rU_z^\dagger(-p)\,. \qquad (6)$$

This channel is usually called a Pauli channel.[6, 16] Figure 7 shows the result. The lower traces are the χ-matrix for *z*=0, 3, 6, and 11 mm, which are in good agreement with the theoretical χ-matrix:

$$c = \begin{pmatrix} 1-p-q-r & 0 & \multicolumn{2}{c}{\mathbf{0}} \\ 0 & p & & \\ \multicolumn{2}{c}{\mathbf{0}} & q & 0 \\ & & 0 & r \end{pmatrix}, \qquad (7)$$

associated with various values of $p$, $q$, and $r$. Finally, let us consider the other scenario in which one wishes to implement a single-qubit gate described by the unitary matrix $U_{\mathbf{e}}(\theta)$, but which cannot be realized as a perfect gate. There are several reasons why such a situation happens. For example, any qubit-environment interaction may introduce noise into the dynamics of the qubit. Additionally, if the parameters that designate the unitary operation such as rotation angle or rotation axis fluctuate, the operation is imperfect. Suppose that the target unitary operation is a Hadamard operation, we have an imperfect Hadamard gate channel represented in the operator sum representation by

$$\mathcal{E}(r) = \frac{1+|s|}{2} U_x\left(\frac{p}{2}\right) r\, U_x^{\dagger}\left(\frac{p}{2}\right) + \frac{1-|s|}{2} U_x\left(-\frac{p}{2}\right) r\, U_x^{\dagger}\left(-\frac{p}{2}\right). \tag{8}$$

This channel is called an imperfect Hadamard gate channel and can be realized with the same setup as the first experiment but if a different focusing position on the plane of the depolarizer is chosen. Figure 8 shows the result representing this situation. The lower traces are the $\chi$-matrix for z=0, 2, 5, and 8 mm, which are in good agreement with the theoretical $\chi$-matrix:

$$c = \frac{1}{2}\begin{pmatrix} 1 & -i|s| & \multicolumn{2}{c}{0} \\ i|s| & 1 & & \\ \multicolumn{2}{c}{0} & \multicolumn{2}{c}{0} \end{pmatrix}, \tag{9}$$

associated with various values of $s$. To the author's knowledge, this is the first demonstration of a complete characterization of a quantum channel implemented by photonic technologies.

The experimental result shows the feasibility of process tomography as a tool to diagnose a quantum device. It also suggests the potential usefulness of a photonic implementation of a qubit in the study of quantum technology. We can construct a variety of quantum devices by using entangled photon sources, passive linear optical elements, and photodetectors.[18, 19] Even conclusive quantum gates can be implemented using only linear optical elements and photodetectors. By using these techniques, it is, in principle, possible to design a variety of quantum computer or quantum mechanical circuits that can perform special tasks. It will be interesting to characterize such a quantum mechanical circuit by using the present technique and to check experimentally the correctness of the theoretical assumptions made in the design of the circuit. For this purpose, we need a characterization technique for higher dimensional gates, e.g., a two-qubit quantum gate. Theoretical extension of process tomography for two-qubit systems is straightforward and has already been given in the literature.[4, 5] Experimental work in this direction is now in progress in our laboratory. Currently, we are interested in characterization of nontrivial two-qubit channels, such as the classically correlated decoherence channel[20] and the conditional quantum phase gate channel.[21] Characterization of two-qubit photonic channels awaits future studies.

## 4. CONCLUSION

We have demonstrated characterization of quantum states and quantum channels. Both the quantum states and the quantum channels were reconstructed from a set of experimental data obtained from tomographic measurements of many identically prepared copies of the system. We experimentally characterized the quantum states of a photonic two-qubit system and photonic single-qubit quantum channels. Our tomographic technique will be of practical importance to the development of various photonic quantum devices and circuits in the future.

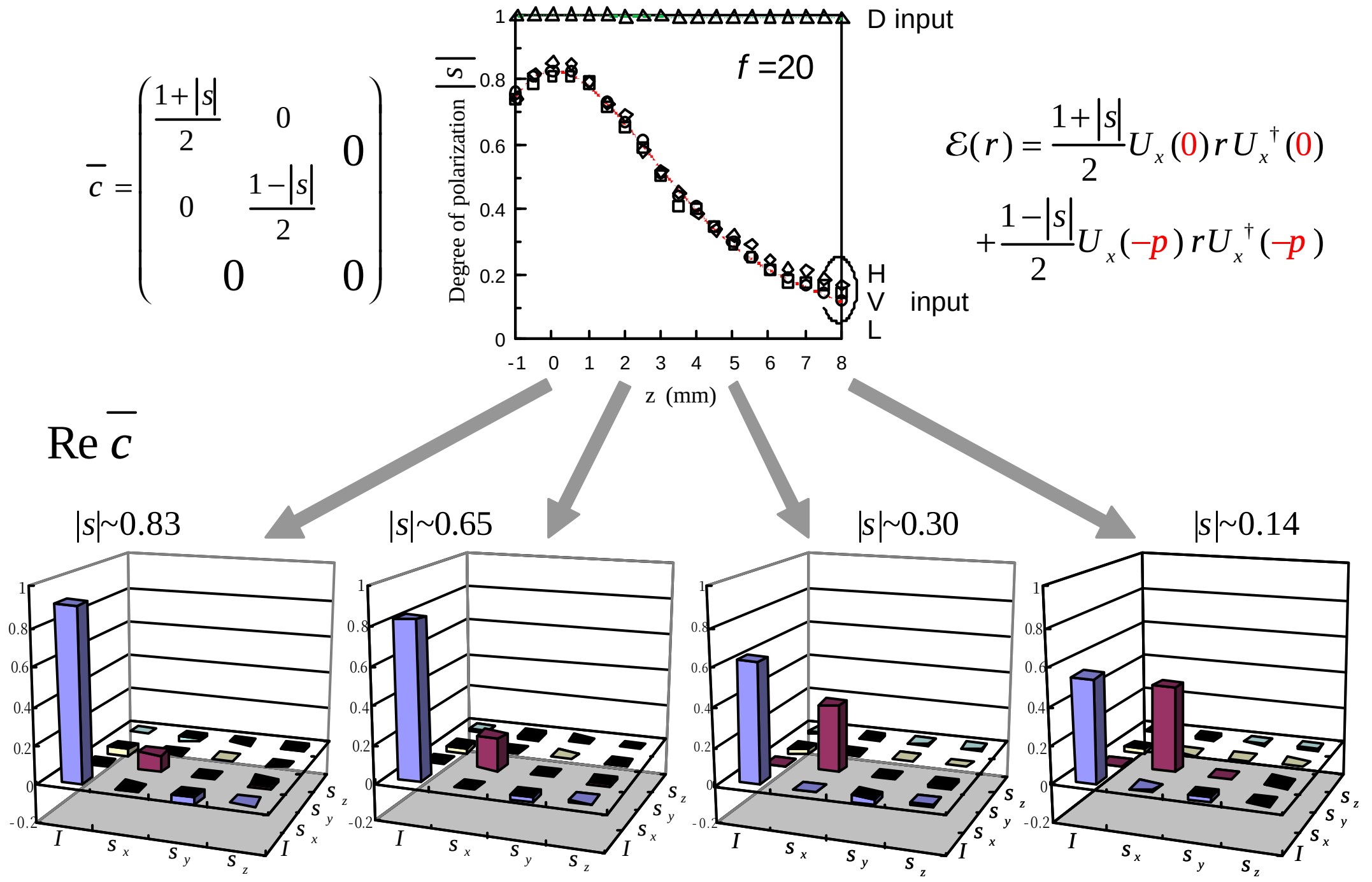

**Figure 6** Experimental result for the partially bit-flip channel. The upper trace indicates the degree of polarization of the photon at the detector as a function of the relative position $z$ of the two lenses and the depolarizer for four input polarization states. The lower traces are the evaluated $\chi$-matrix for $z$=0, 2, 5, and 8 mm. Only the real parts of the matrix elements are shown (the imaginary parts are negligible).

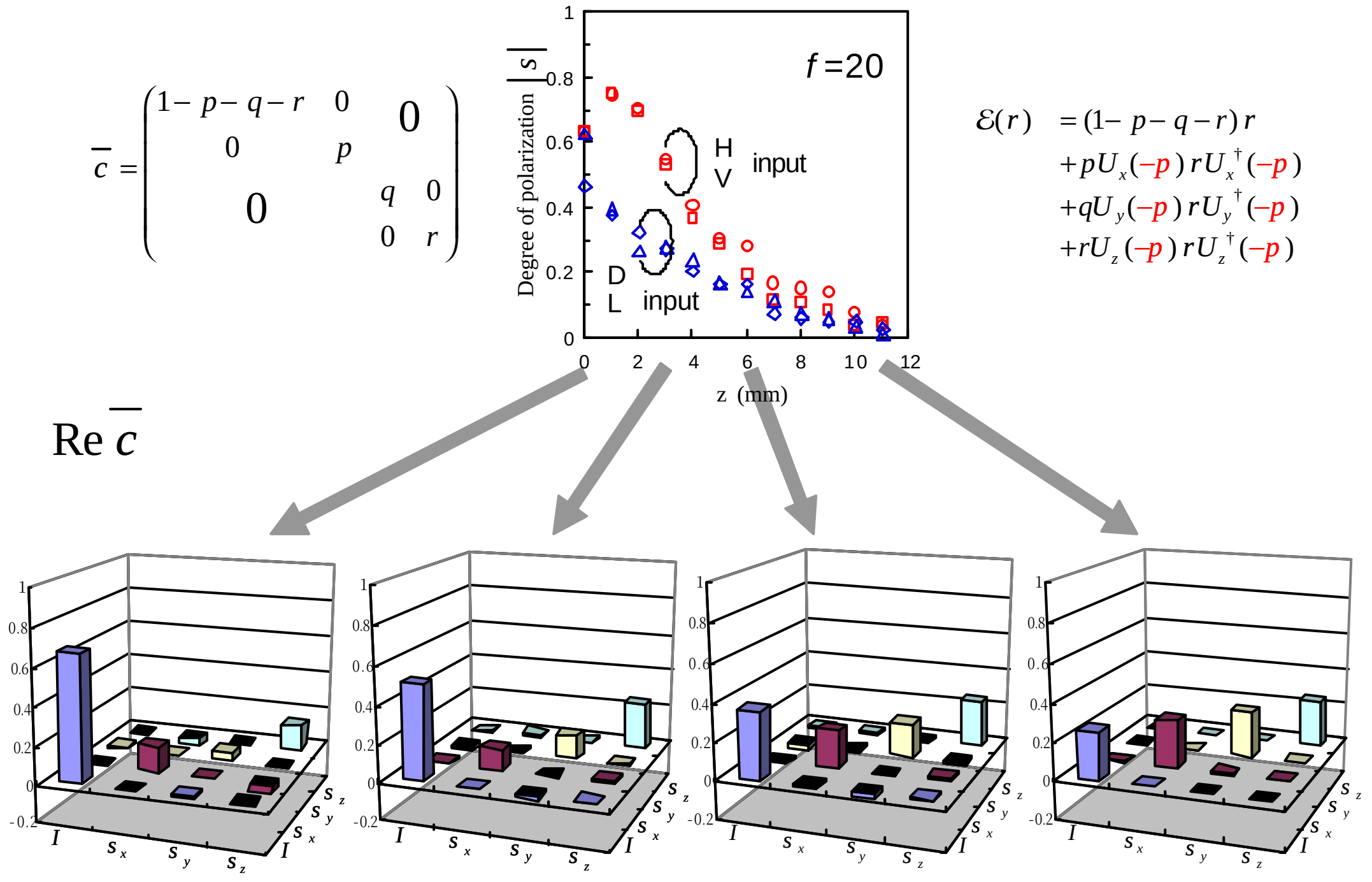

**Figure 7** Experimental result for the Pauli channel. The lower traces are the evaluated $\chi$-matrix for $z$=0, 3, 6, and 11 mm. Only the real parts of the matrix elements are shown (the imaginary parts are negligible).

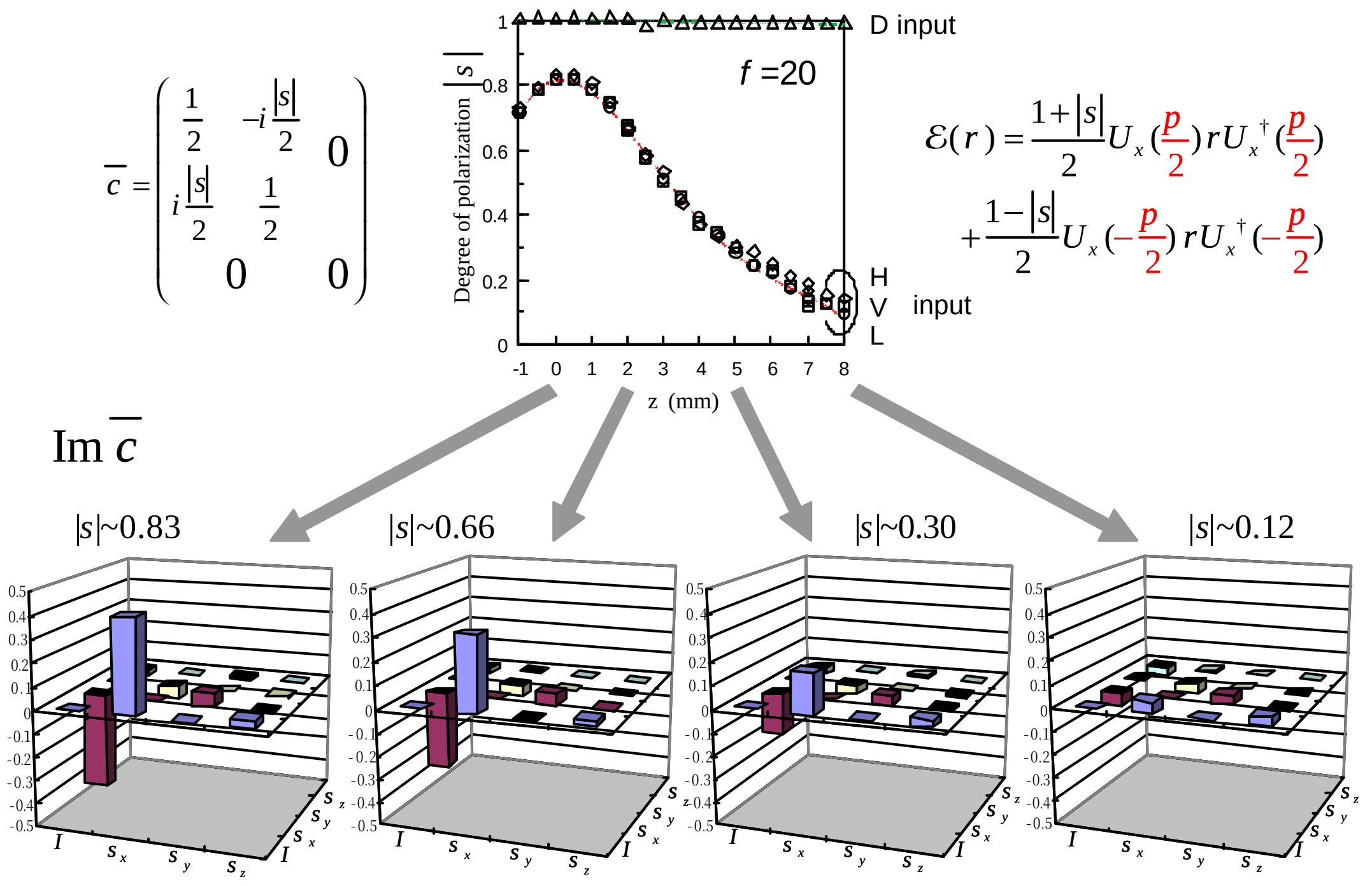

**Figure 8** Experimental result for the imperfect Hadamard gate channel. The lower traces are the evaluated $\chi$-matrix for $z$=0, 2, 5, and 8 mm. Only the imaginary parts of the matrix elements are shown (the real parts are consistent with the model).